\documentclass[12pt]{article}
\usepackage{epsfig,picinpar,floatflt,amssymb}
\title{Form Factors in Off--Critical Superconformal 
Models}

\author{G. Mussardo
  \footnote{Dipartimento di Fisica, 
Universita' dell'Insubria, Como (Italy). 
Supported by EC TMR Programme {\em Integrability, non--perturbative
effects and symmetry in QFT}, grant FMRX-CT96-0012.}}

\begin{document}

\maketitle
              
\abstract{
We discuss the determination of the lowest Form Factors relative to the
trace operators of $N=1$ Super Sinh-Gordon Model. 
Analytic continuations of these Form Factors as functions of the coupling 
constant allows us to study a series of models in a uniform way,
among  these the latest model of the Roaming Series and a class of minimal
supersymmetric models.}
\newcommand{\EQ}{\begin{equation}}
\newcommand{\EN}{\end{equation}}
\newcommand{\hs}{\hspace{0.1cm}}
\newcommand{\spz}{\hspace{0.7cm}}
\newcommand{\goto}{\rightarrow}
\newcommand{\be}{\beta}
\newcommand{\zb}{\bar{z}}
\newcommand{\barr}{\overline}

\vspace{5mm}

\noindent
{\large{\bf Introduction}}

A full control of a Quantum Field Theory (QFT) is reached once 
the complete set of correlation functions of its fields is known. 
Perturbation theory often proves to be an inadequate approach to 
this problem therefore more effective and powerful methods need to be
developed. In this respect, one of the most promising methods is the 
Form Factor approach which is applicable to integrable models
\cite{KW,Smirnov}. This consists in computing exactly all matrix elements
of the quantum fields and then using them to obtain the spectral 
representations of the correlators. In addition to the rich and interesting
mathematical structure presented by the Form Factors themselves (which has 
been investigated in a series of papers, among  which [2-11], 
the resulting spectral series usually show a remarkable 
convergent behaviour which allows approximation of the correlators (or 
quantities related to them) within any desired accuracy [11-17]. 

In this talk, based on the original paper \cite{GMsusy}, I will briefly
discuss some new results obtained for those QFT which are invariant under a
supersymmetry transformation which mixes the elementary bosonic and 
fermionic excitations. Namely, I will present 
the result relative to the lowest Form Factors of the Super Sinh-Gordon
Theory  (SShG) and of those models which can be obtained by its analytic 
continuations. In these models, the degeneracy of the spectrum dictated 
by the supersymmetry implies the existence of multi--channel scattering 
processes and the resulting $S$-matrix is necessarily non--diagonal. In
this case, the complete determination of the matrix elements of the quantum
fields  for an arbitrary number of asymptotic particles is still 
an open problem. 
   
\noindent
{\large{\bf The SShG and Superconformal Models}}

Scattering theories of integrable super-symmetric theories have 
been discussed in detail in Schoutens's paper  
\cite{Schoutens} (see also \cite{HM}). We refer to these papers and 
also to \cite{GMsusy} for a general discussion of the features of 
these theories and the relative notation. Let us concentrate then our 
attention to the main subjects of this talk, i.e. the Super Sinh-Gordon model 
and the integrable deformations of the superconformal models. 
In the euclidean space, the Super Sinh-Gordon model can be
defined in terms of its action 
\EQ
{\cal A} = \int d^2z d^2\theta \left[
\frac{1}{2} D\Phi \barr D \Phi + i \frac{m}{\lambda^2} 
\cosh\lambda\Phi \right] \,\,\, ,
\label{actionSSHG}
\EN 
where the covariant derivatives are defined as 
$
D = \partial_{\theta} - \theta \,\partial_z$; $
\barr D = \partial_{\barr\theta} -\barr\theta \,\partial_{\bar z}$,
and the superfield $\Phi(z,\bar z,\theta,\barr\theta)$ has an expansion as 
\EQ
\Phi(z,\bar z,\theta,\barr\theta)= \varphi(z,\bar z) + 
\theta \psi(z,\bar z) + \barr \theta \,\barr \psi(z,\bar z) + 
\theta \barr \theta \,{\cal F}(z,\bar z) \,\,\,.
\label{superfield}
\EN 
The integration on the $\theta$ variables and also the elimination
of the auxiliary field ${\cal F}(z,\bar z)$ by means of its algebraic equation 
of motion leads to the Lagrangian 
\EQ
{\cal L} = \frac{1}{2} 
(\partial_z\varphi \partial_{\bar z}\varphi + 
\barr\psi \partial_z \barr\psi + \psi\partial_{\bar z} \psi ) 
+\frac{m^2}{2 \lambda^2} \sinh^2\lambda\varphi + 
i m \barr\psi \psi \, \cosh\lambda\varphi \,\,\, .
\label{lagrangianSSHG}
\EN 

Of all the different ways of looking at the SSHG model, one of 
the most convenient is to consider it as a deformation of the 
superconformal model described by the action 
\EQ
{\cal A_0} = \frac{1}{2}\int(  
\partial_z\varphi \partial_{\bar z}\varphi + 
\barr\psi \partial_z \barr\psi + \psi\partial_{\bar z} \psi) \,\,\,.
\label{SCFT3/2}
\EN
This superconformal model has central charge $C=3/2$. At this point, 
it is useful to briefly remind some properties of the superconformal 
models and their deformations.  

For a generic superconformal model, the supersymmetric 
charges can be represented by the differential operators 
$
Q = \partial_{\theta} + \theta \,\partial_z$; $ 
\overline Q = \partial_{\bar\theta} +\barr\theta \,\partial_{\bar z}$. 
The analytic part of the stress-energy tensor $T(z)$ and the 
current $G(z)$ which generates the supersymmetry combine 
themselves into the analytic superfield 
$ 
W(z,\theta) = G(z) + \theta T(z) \,\,\, ,
$, 
which is called the super stress-energy tensor. 
For the anti-analytic sector we have correspondingly 
$\barr W(\bar z) = \barr G(\bar z) +\barr\theta \,\barr T(\bar z)$. 
These fields are mapped one into the other by means of the 
super-charges. 
As it is well known \cite{superconformal}, reducible unitary
representations of the $N=1$ superconformal symmetry occurs 
for the discrete values of the central charge 
\EQ
C = \frac{3}{2} -\frac{12}{m (m+2)} \,\,\,. 
\label{discretec}
\EN 
At these values, realizations of the $N=1$ superconformal algebra 
are given in terms of a finite number of superfields in the Neveu-Scwartz 
sector and a finite number of ordinary conformal primary fields 
in the Ramond sector. Their conformal dimensions are given by 
\EQ
\Delta_{p,q} = \frac{[(m+2) p -m q]^2-4}{8 m (m+2)} +
\frac{1}{32} [1 -(-1)^{p-q}] \,\,\, ,
\label{anomalousdimensions}
\EN 
where $p-q$ even corresponds to the primary Neveu-Schwartz 
superfields ${\cal N}_{p,q}^{(m)}(z,\theta)$ and $p-q$ odd to the 
primary Ramond fields $R_{p,q}^{(m)}(z)$. These fields enter the 
so-called superconformal minimal models ${\cal SM}_m$.  

The Witten index $Tr (-1)^F$ of the superconformal models can be 
computed by initially defining them on a cylinder \cite{KMS}: the 
Hamiltonian on the cylinder is given by $H=Q^2=L_0-C/24$, where 
as usual $C/24$ is the Casimir energy on the cylinder and 
$L_0 = \frac{1}{2\pi i}\oint dz z T(z)$. Considering that for any
conformal state $\mid a\rangle$ with $\Delta > C/24$ there is the 
companion state $Q\mid a\rangle$ of opposite fermionic parity, their 
contributions cancel each other in $Tr (-1)^F$ and therefore 
only the ground states with $\Delta = C/24$ (which are not 
necessarily paired) enter the final expression of the Witten 
index. For the minimal models, there is a non-zero 
Witten index only for $m$ even. Therefore the lowest superconformal 
minimal model with a non-zero Witten index is the one with 
$m=4$, which has a central charge $C=1$ and corresponds to 
the class of universality of the critical Ashkin-Teller model. 
The superconformal theory with $C=3/2$ made of free 
bosonic and fermionic fields also has a non-zero Witten index, 
because an unpaired Ramond field $R(z)$ is explicitly given by the spin 
field $\sigma(z)$ of Majorana fermion $\psi(z)$ with conformal 
dimension $\Delta =1/16$, i.e. by the magnetization operator of 
the Ising model.  

The above observations become important in the understanding 
the off-critical dynamics relative to the deformation of 
the action of the superconformal minimal models ${\cal SM}_m$ by 
means of the relevant supersymmetric Neveu-Schwartz operator 
${\cal N}_{1,3}^{(m)}(z,\theta)$. In fact, as shown in \cite{KMS}, 
the massless Renormalization Group flow generated by such an operator 
preserves the Witten index. Therefore the long distance 
behaviour of the deformed ${\cal SM}_m$ minimal model -- controlled 
by the action 
$ 
{\cal A}_m + \gamma \int d^2z \,d^2\theta \,{\cal N}_{1,3}^{(m)}$ 
(for a positive value of the coupling constant $\gamma$, with 
the usual conformal normalization of the superfield) -- is ruled by 
the fixed point of the minimal model ${\cal SM}_{m-2}$. Therefore 
this action describes the RG flow 
$ 
{\cal SM}_{m} \rightarrow {\cal SM}_{m-2} 
$, with a corresponding jump in steps of two of the central 
charge\footnote{This is in contrast of what happens in 
the ordinary conformal minimal models, where the deformation 
of the conformal action by means of the operator $\phi_{1,3}$ induces 
a massless flow between two next neighborod minimal models.}, i.e. 
$\Delta C =C(m)-C(m-2)$.  
Therefore, a cascade of massless flows which start  
from $C=3/2$ and progress by all ${\cal N}_{1,3}^{(m)}$
deformations of ${\cal SM}_m$ met along the way must  
necessarily pass through the model with $C=1$ in the second to last 
step rather than the lowest model\footnote{As well known, the lowest 
model with $C=7/10$ corresponds to the class of universality of the 
Tricritical Ising Model.} with $C=7/10$. 
This is indeed the scenario which is described by a specific analytical
continuation of the coupling constant of the Super-Sinh-Gordon model, 
the so-called Roaming Models. 

Let us now discuss the deformation of 
the $C=3/2$ superconformal theory which leads to 
the SShG model. At the conformal point, the explicit realization 
of the component of the super stress-energy tensor are given by 
\EQ
\begin{array}{l}
T(z) = -\frac{1}{2}\left[
(\partial_z\varphi)^2 - \psi \partial\psi\right] \,\,\, ;\\
G(z) = \,i\psi \partial_z \varphi \,\,\, ,
\end{array}
\label{superstressc3/2}
\EN 
and they satisfy the conservation laws 
$\partial_{\bar z} T(z) = \partial_{\bar z} G(z) =0$. 
Once this superconformal model is deformed according to the Lagrangian 
(\ref{lagrangianSSHG}), the new conservation laws are given by 
$ 
\partial_{\bar z} T(z,\bar z) =\partial_z \Theta(z,\bar z)
$; $ 
\partial_{\bar z} G(z,\bar z) = \partial_z \chi(z,\bar z)$, 
where 
\begin{eqnarray}
&& \Theta(z,\bar z) = \frac{m^2}{2 \lambda^2} \sinh^2\lambda\varphi + 
i m \barr\psi \psi \, \cosh\lambda\varphi \,\,\, ,
\label{tracethetag}\\
&& \chi(z,\bar z) = \,\frac{m}{\lambda}\barr \psi \,\sinh\lambda\varphi
\,\,\, .\nonumber 
\end{eqnarray}
For the anti-analytic part of the super stress-energy tensor
we have 
$ 
\partial_{z} \bar T(z,\bar z) =\partial_{\bar z} \Theta(z,\bar z)
$; $ 
\partial_{z} \barr G(z,\bar z) = \partial_{\bar z}\barr \chi(z,\bar z) 
$, 
where $\Theta(z,\bar z)$ is as before and the other fields are given 
by 
\begin{eqnarray}
&& \barr G(z,\bar z) = -i \barr \psi \partial_{\bar z} \varphi \,\,\,;
\label{tracethetag2} \\
&& \barr\chi(z,\bar z) = \,\,\frac{m}{\lambda}\psi \,\sinh\lambda\varphi
\,\,\, .\nonumber 
\end{eqnarray}
The operators $\Theta(z,\bar z)$, $\chi(z,\bar z)$ and 
$\barr\chi(z,\bar z)$ belong to the trace of the supersymmetric 
stress-energy tensor and they are related each other by 
\EQ
\begin{array}{lll}
\Theta(z,\bar z) = \{\chi(z,\bar z),{\cal Q}\} &,& 
\partial_z \chi(z,\bar z) = [\Theta(z,\bar z),{\cal Q}]\,\,\,; \\
\Theta(z,\bar z) = \{\barr \chi(z,\bar z),\barr {\cal Q}\} &,& 
\partial_{\bar z} \barr \chi(z,\bar z) =[\Theta(z,\bar z),\barr 
{\cal Q}] \,\,\, ,
\end{array}
\label{susytransftrace}
\EN
where the charges of supersymmetry are expressed by 
\EQ
\begin{array}{l}
Q = \int G(z,\bar z) dz + \chi(z,\bar z) d\bar z\,\,\, ;\\
\barr Q = \int \barr G(z,\bar z) d\bar z + \bar \chi(z,\bar z) dz 
\,\,\, .
\end{array}
\label{integralcharges}
\EN 
In addition to the above conservation laws, the SShG model possesses 
higher integrals of motion which were explicitly determined in 
\cite{FGS}. Therefore its scattering processes are purely elastic 
and factorizable. and its two-body $S$-matrix is discussed in the 
next section. 

\noindent
{\large{\bf The $S$-matrix of the SSHG}}

The exact $S$-matrix of the SSHG model has been determined in 
\cite{Ahn}. It is given by 
\EQ 
S(\beta) = Y(\beta)\,
\left(\begin{array}{cccc} 
1-\frac{2 i \sin\pi\alpha}{\sinh\beta} & 
\frac{-\sin\pi\alpha}{\cosh\frac{\beta}{2}} 
& 0 & 0 \\ 
\frac{-\sin\pi\alpha}{\cosh\frac{\beta}{2}} & 
-1 -\frac{2 i \sin\pi\alpha}{\sinh\beta} & 0 & 0 \\
0 & 0 &-\frac{i\sin\pi\alpha}{\sinh\frac{\beta}{2}} & 1 \\
0 & 0 & 1 & 
-\frac{i\sin\pi\alpha}{\sinh\frac{\beta}{2}}
\end{array}
\right) 
\label{SmatSSHG}\\
\EN 
where 
\EQ
Y(\beta) = \frac{\sinh\frac{\beta}{2}}{\sinh\frac{\beta}{2} + 
i \sin\pi\alpha} \, U(\beta,\alpha) \,\,\, ,
\EN 
and the function $U(\beta)$ is given by 
\EQ
U(\beta) = \exp\left[i 
\int_0^{\infty} \frac{dt}{t} 
\frac{\sinh\alpha t\,\sinh(1-\alpha)t}{\cosh^2\frac{t}{2} \cosh t} 
\sin\frac{\beta t}{\pi}\right] \,\,\,.
\label{integralU}
\EN 
The angle $\alpha$ is a positive quantity, related to the coupling
constant $\lambda$ of the model by 
\EQ
\alpha = \frac{1}{4 \pi} \frac{\lambda^2}{1 + \frac{\lambda^2}{4 \pi}}
\,\,\, .
\label{relationcouplings}
\EN 
This equation implies that the SShG is a quantum field theory 
invariant under the strong-weak duality 
$ 
\lambda \rightarrow \frac{4\pi}{\lambda} 
$. It is now interesting to analyse several analytic continuations 
of the above $S$-matrix in the parameter $\alpha$. 

Under the analytic continuation $\alpha \rightarrow -\alpha$, the SShG 
model goes into the Super Sine-Gordon model and the $S$-matrix 
(\ref{SmatSSHG}) describes in this case the scattering of the lowest 
breather states of the latter model. At the particular value 
$\alpha=\pi/3$ we can consistently truncate the theory at the Super 
Sine-Gordon breather sector only\footnote{The general class of these models 
is obtained by taking 
$\alpha = \pi/(2 N +1)$, $N=1,2,\ldots$ and they correspond to 
the supersymmetric deformation of the non-unitary minimal superconformal 
models with central charge $C=-3N (4N+3)/(2N+2)$.}. At this value, 
the pole at $\beta =2 \pi i/3$ of the $S$-matrix can be regarded 
as due to the bosonic and fermionic one-particle states $\mid
b(\beta)\rangle $ and $\mid f(\beta)\rangle$. These particles are 
therefore bound states of themselves, in the channels 
$ b b \rightarrow b \rightarrow bb$, $ f f \rightarrow b \rightarrow f f 
$, for the bosonic particle $b$, and in the channels 
$ b f \rightarrow f \rightarrow f b$, $f b \rightarrow f \rightarrow b f
$, for the fermionic particle $f$. There is of course a price to pay for 
this truncation: this means that the residues of the $S$-matrix at 
these poles will be purely imaginary. Such a model 
therefore would be the supersymmetric analogous of the Yang-Lee model 
for the ordinary Sine-Gordon model \cite{YL}. It was considered 
originally by Schoutens \cite{Schoutens} and it has been identified 
with the off-critical supersymmetric deformation of the non-unitary 
superconformal minimal model with central charge $C=-21/4$. 

Another analytic continuation of the $S$-matrix gives rise to the 
so--called {\em Roaming Models}. Notice, in fact, that  
the $S$ matrix (\ref{SmatSSHG}) of the SShG model has zeros in 
the physical strip located 
at $\alpha_1=i\pi\alpha$ and $\alpha_2=i\pi(1-\alpha)$. By varying the 
coupling constant $\lambda$, they move along the imaginary axis and 
they finally meet at the point $i\pi/2$, at the self-dual value of the 
coupling constant $\lambda^2 = 4\pi$. If we further increase the 
value of the coupling constant, they simply swap positions. 
But there is a more interesting possibility: as first proposed by 
Al. Zamolodchikov for the analogous case of the ordinary Sinh-Gordon 
model \cite{roaming}, once the two zeros meet at $i\pi$, they can  
enter the physical strip by taking complex values of the 
coupling constant. In this way, the location of the two 
zeros are given by 
$ 
\alpha_{\pm} = \frac{1}{2} \pm i \alpha_0$. 
From the analytic $S$-matrix theory, the existence of complex zeros 
in the physical strip implies the presence of resonances 
in the system. By analysing the finite-size behaviour of the theory by
means of the Thermodynamic Bethe Ansatz \cite{Martins}, the interesting 
result is that the net effect of these resonances consists in an infinite 
cascade of massless Renormalization Group flows generated by the 
Neveu-Schwartz fields ${\cal N}_{1,3}^m$ and passing through all 
minimal superconformal model ${\cal SM}_m$ with non-zero Witten index. 
As previously discussed, the ending
point of this infinite-nested RG flow should describe the ${\cal N}_{1,3}$
deformation of the superconformal model ${\cal SM}_{4}$. Is this really 
the case? By taking the limit $\alpha \rightarrow i \infty$ into the
$S$-matrix (\ref{SmatSSHG}), it is easy to see that it reduces 
the $S$-matrix of the Sine-Gordon model at $\xi=2\pi$. Since 
this $S$-matrix describes a massive deformation of the $C=1$ model, 
in order to confirm the above roaming trajectory scenario 
the only thing that remains to check is the comparison of the anomalous 
dimension of deforming field. In the Sine-Gordon model at $\xi=2\pi$, 
the anomalous dimension of the deforming field is  
$\Delta =2/3$, which is indeed the conformal dimension of the top 
component of the superfield ${\cal N}_{1,3}$ in the model ${\cal SM}_4$.

\noindent
{\large {\bf Form Factors of the Trace Operators of the SShG Model}}

For integrable quantum field theories, the knowledge of the
$S$-matrix is very often the starting point for a complete 
solution of quantum field dynamics in terms of an explicit
construction of the correlation functions of all fields of 
the theory. This result can be obtained by computing first 
the matrix elements of the operators on the asymptotic 
states (the so-called Form Factors) \cite{KW,Smirnov} and then 
inserting them into the the spectral representation of the 
correlators. For instance, in the case of the two-point 
correlation function of a generic operator ${\cal O}(z,\bar z)$ 
we have 
\EQ
{\cal G}(z,\bar z) = 
\langle 0 \mid {\cal O}(z,\bar z) {\cal O}(0,0)\mid 0\rangle = 
\int_0^{\infty} da^2 \,\rho(a^2) \, K_0(a \sqrt{z\bar z}) \,\,\, ,
\label{spectralrep}
\EN 
where $K_0(x)$ is the usual Bessel function. The spectral density 
$\rho(a^2)$ is given in this case by 
\begin{eqnarray}
\rho(a^2) &=& \sum_{n=0}^{\infty} 
\int 
\frac{d\beta_1}{2\pi}
\cdots 
\frac{d\beta_n}{2\pi}
\,\delta(a-\sum_i^n m \cosh\beta_i) \,\,\delta(\sum_i^n \sinh\beta_i) 
\times \label{rho} \\
&& \,\,\,\,\,\,\,\,\,\,\,\,\,\,
\mid \langle 0\mid {\cal O}(0,0)\mid A_1(\beta_1)\ldots
A_n(\beta_n) \rangle\mid^2 \,\,\, . \nonumber 
\end{eqnarray} 
The Form-Factor approach has proved to be extremely successful 
for theories with scalar $S$-matrix, leading to an explicit 
solution of models of statistical mechanics interest such as the 
Ising model \cite{KW,JLG,YZ}, the Yang-Lee model \cite{ZamYL} 
or quantum field theories defined by a lagrangian, like 
the Sinh-Gordon model \cite{KM,FMS}. On the contrary, 
for theories with a non-scalar $S$-matrix the functional
equations satisfied by the Form Factors are generally quite 
difficult to tackle and a part from the Sine-Gordon model or theories 
which can be brought back to it \cite{KW,Smirnov,KF}, there is presently no 
mathematical technique available for solving them in their 
full generality. Also in this case, however, the situation is not 
as impractible as it might seem at first sight. The reason 
consists in the fast convergent behaviour of the spectral 
representation series which approximates the correlation functions with a 
high level of accuracy even if truncated at the first available 
matrix elements [11--17]. 
In the light of this fact, we will compute 
the lowest matrix elements of two of the most important operators 
of the theory, namely the trace
operators  $\Theta(z,\bar z)$, $\chi(z,\bar z)$ and 
$\barr\chi(z,\bar z)$ of the supersymmetric stress-energy tensor 
of the SShG model. For the operator $\Theta(0,0)$, they 
are given by\footnote{They depend on the difference of rapidities 
$\beta =\beta_1-\beta_2$ since $\Theta(0,0)$ is a scalar operator.} 
\EQ
\begin{array}{l}
F_{bb}^{\Theta}(\beta) = \langle 0\mid \Theta(0,0)
\mid b(\beta_1) b(\beta_2)\rangle \,\,\,;\\
F_{ff}^{\Theta}(\beta) 
=\langle 0\mid \Theta(0) \mid f(\beta_1) f(\beta_2)\rangle \,\,\, ,
\end{array}
\label{FFTheta}
\EN 
whereas for the operators $\chi(0,0)$ we have instead
\EQ
\begin{array}{l} 
F_{bf}^{\chi}(\beta_1,\beta_2) = \langle 0\mid \chi(0,0)
\mid b(\beta_1) f(\beta_2)\rangle \,\,\,;\\
F_{fb}^{\chi}(\beta_1,\beta_2) = \langle 0\mid \chi(0,0)
\mid f(\beta_1) b(\beta_2)\rangle \,\,\,,
\end{array}
\label{FFchi}
\EN
(with an analogous result for the lowest Form Factors of the 
operator $\barr\chi(0,0)$). Since the operators $\Theta$, $\chi$ 
and $\barr\chi$ are related each other by supersymmetry, as 
consequence of eqs.\,(\ref{susytransftrace}) we have 
\begin{eqnarray}
&& F_{bb}^{\Theta}(\beta) = \,\,\,\omega \,\left(e^{\beta_1/2} \,
F_{fb}^{\chi} + e^{\beta_2/2}\,F_{bf}^{\chi}\right) \,\,\,;
\nonumber \\
&& F_{ff}^{\Theta}(\beta) = -\barr\omega \,\left(e^{\beta_2/2} \,
F_{fb}^{\chi} - e^{\beta_1/2}\,F_{bf}^{\chi}\right) \,\,\,;
\\
&& F_{bb}^{\Theta}(\beta) = \,\,\,\barr\omega \,\left(e^{-\beta_1/2} \,
F_{fb}^{\barr\chi} + e^{-\beta_2/2}\,F_{bf}^{\barr\chi}\right) \,\,\,;
\nonumber \\
&& F_{ff}^{\Theta}(\beta) = -\omega \,\left(e^{-\beta_2/2} \,
F_{fb}^{\barr\chi} - e^{-\beta_1/2}\,F_{bf}^{\barr\chi}\right) \,\,\,.
\nonumber 
\end{eqnarray}
It is therefore sufficient to compute the two-particle Form Factors 
of the operator $\Theta(z,\bar z)$ for determining those of 
$\chi(z,\bar z)$ and $\barr\chi(z,\bar z)$. Let us discuss the functional 
equations satisfied by $F_{bb}^{\Theta}(\beta)$ and 
$F_{ff}^{\Theta}(\beta)$.
The first set of equations (called the unitarity equations) rules the 
monodromy properties of the matrix elements as dictated by the 
$S$-matrix amplitudes  
\begin{eqnarray}
&& F_{bb}^{\Theta}(\beta) = S_{bb}^{bb}(\beta)\,
F_{bb}^{\Theta}(-\beta) + S_{bb}^{ff}(\beta) \,
F_{ff}^{\Theta}(-\beta) \,\,\,;
\label{unitarityFF} \\
&& F_{ff}^{\Theta}(\beta) = S_{ff}^{bb}(\beta)\,
F_{bb}^{\Theta}(-\beta) + S_{ff}^{ff}(\beta) \,
F_{ff}^{\Theta}(-\beta) \,\,\,,\nonumber
\end{eqnarray}
where $S_{bb}^{bb}(\beta)$ is the scattering amplitude of two 
bosons into two bosons and similarly for the other amplitudes. 
The second set of equations (called crossing equations) 
express the locality of the operator $\Theta(z,\bar z)$ 
\EQ
\begin{array}{l}
F_{bb}^{\Theta}(\beta + 2\pi i) = 
F_{bb}^{\Theta}(-\beta) \,\,\,;\\
F_{ff}^{\Theta}(\beta + 2\pi i) = 
F_{ff}^{\Theta}(-\beta) \,\,\,.
\end{array}
\label{crossingFF}
\EN 

The way to solve these equations was found in \cite{GMsusy} and passes 
by the definition of two auxiliary functions $H_{\pm}(\beta)$ which 
are related to the scattering amplitudes the pure fermionic 
sector of the theory. They can be used as building blocks for 
constructing the matrix elements $F_{bb}^{\Theta}(\beta)$ and 
$F_{ff}^{\Theta}(\beta)$. For lacking of space, we give here only 
their final expressions, 
\begin{eqnarray}
&&F_{bb}^{\Theta}(\beta) = 2\pi m^2\,\frac{\tilde F_{bb}(\beta)}
{\tilde F_{bb}(i\pi)} \,\,\,;
\label{finalFF} \\
&&F_{ff}^{\Theta}(\beta) = 2\pi m^2\, 
\frac{\tilde F_{ff}(\beta)}{\tilde F_{ff}(i\pi)} \,\,\, ,\nonumber
\end{eqnarray}
where 
\begin{eqnarray}
&& \tilde F_{bb}(\beta) = \left[
\cosh^2\frac{\beta}{4} \,H_+(\beta) - 
\sinh^2\frac{\beta}{4} \,H_-(\beta)\right]\,G(\beta) \,\,\, ,
\label{finaltildeFF} \\
&& \tilde F_{ff}(\beta) =  
\sinh\frac{\beta}{2} \,
\left[H_+(\beta) + H_-(\beta)\right]\,G(\beta) \,\,\, .  \nonumber 
\end{eqnarray}
The functions $G(\beta)$ and $H_{\pm}(\beta)$ are given by their integral
representation   
\EQ
G(\beta) = 
\exp\left[- \int_0^{\infty} \frac{dt}{t} 
\frac{\sinh\alpha t\,\sinh(1-\alpha)t}{\cosh^2\frac{t}{2} \sinh t \cosh t} 
\sin^2\frac{(i \pi -\beta) t}{2 \pi}\right] \,\,\,.
\label{integralG}
\EN 
\begin{eqnarray}
&& H_+(\beta) = \exp\left[4 \int_0^{\infty} \frac{dt}{t} 
\frac{\sinh\alpha t \,\sinh (1-\alpha)t}{\cosh t \sinh 2t} 
\sin^2\left(\frac{\beta - 2\pi i}{2\pi}\right)t \right] \,\,\,;
\label{H+-} \\
&& H_-(\beta) = 
\exp\left[4 \int_0^{\infty} \frac{dt}{t} 
\frac{\sinh\alpha t \,\sinh (1-\alpha)t}{\cosh t \sinh 2t} 
\sin^2\frac{\beta}{2\pi}t \right] \,\,\,. \nonumber 
\end{eqnarray}

These matrix elements are normalised as $F_{bb}(i\pi) = F_{ff}(i\pi) 
= 2\pi m^2$. Notice that for large values of $\beta$, $F_{bb}^{\Theta}(\beta)$ 
tends to a constant whereas $F_{ff}^{\Theta}(\beta) \simeq e^{\beta/2}$, 
both behaviour in agreement with Weinberg's power counting theorem 
of the Feynman diagrams. 

The Form Factors (\ref{finalFF}) can now be used   
to estimate the correlation function $C(r)= \langle \Theta(r)
\Theta(0)\rangle$ by means of formulas (\ref{spectralrep}) and (\ref{rho}). 
In the free limit, the correlator is simply expressed in terms of 
Bessel functions, 
\EQ
C(r) = m^4 \left(K_1^2(m r) + K_0^2(m r)\right)\,\,\,. 
\label{freecor}
\EN 
For a finite value of $\alpha$, a numerical integration of 
(\ref{spectralrep}) produces the graphs shown in Figure 1.

\vspace{-3cm}
\begin{center}
\begin{minipage}[b]{.65\linewidth}
\centering\psfig{figure=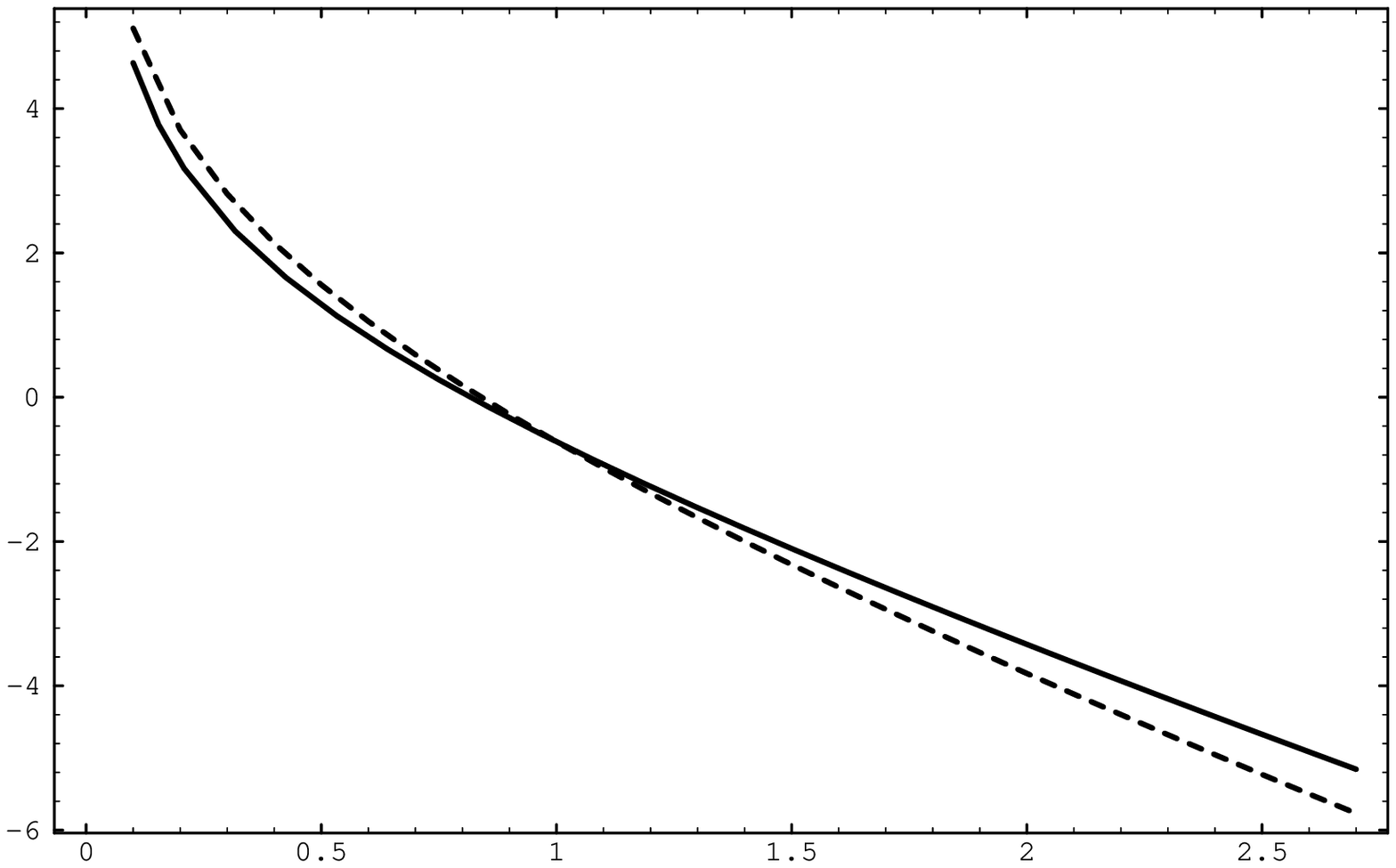,width=\linewidth}

\vspace{-3cm}
\begin{center}
Figure 1. Logarithm of the correlation function $m^{-4} \langle
\Theta(x) \Theta(0) \rangle$ versus $m r$ for $\alpha=0$ (full line) and for 
$\alpha=0.5$ (dashed line). 
\end{center}
\end{minipage} \hfill
\end{center}

As it was expected, in the ultraviolet limit the curve relative to a 
finite value of $\alpha$ is stepest than the curve relative to the 
free case whereas it decreases slower at large values of $mr$.  
This curve is expected to correctly capture the long distance behaviour 
of the correlator and to provide a reasonable estimate of their 
short distance singularity. However, for the exact estimation of the power
law singularity at the origin one would need the knowledge of 
all higher particle Form Factors. 

\noindent 
{\large{\bf $C$-theorem Sum Rule}}

The SShG model can be seen as a massive 
deformation of the superconformal model with the central charge 
$C=\frac{3}{2}$. While this fixed point rules the ultraviolet 
properties of the model, its large distance behaviour is controlled 
by a purely massive theory with $C=0$. The variation of the central 
charge in this RG flow is dictated by the $C$-theorem of Zamolodchikov 
\cite{Zamcth}, which can be formulated in terms 
of a sum rule \cite{Cardycth}
\EQ
\Delta C = \frac{3}{4\pi} \int d^2x\,\mid x\mid^2\, 
\langle 0\mid\Theta(x) \Theta(0)\mid 0\rangle_{conn} \,= 
\int_0^{\infty} d\mu\, c(\mu)\,\, ,
\label{sumrule}
\EN 
where $c(\mu)$ is given by 
\begin{eqnarray}
& &c(\mu) = \frac{6}{\pi^2} \frac{1}{\mu^3} {\rm Im} \,G(p^2=-\mu^2) \,\,\,,
\nonumber \\
&& G(p^2) = \int d^2x e^{-i p x}  
\langle 0\mid\Theta(x) \Theta(0)\mid 0\rangle_{conn} \,\,\,.
\label{spectralcth}
\end{eqnarray}
Inserting a complete set of in-state into (\ref{spectralcth}), the 
spectral function $c(\mu)$ can be expressed as a sum on the FF's 
\begin{eqnarray}
c(\mu) &=& \frac{12}{\mu^3} \sum_{n=1}^{\infty} 
\int \frac{d\beta_1}{2\pi}
\cdots 
\frac{d\beta_n}{2\pi}
\,\delta(\mu -\sum_i^n m \cosh\beta_i) \,\,\delta(\sum_i^n m\sinh\beta_i) 
\times \label{cspect} \\
&& \,\,\,\,\,\,\,\,\,\,\,\,\,\,
\mid \langle 0\mid \Theta(0,0)\mid A_1(\beta_1)\ldots
A_n(\beta_n) \rangle\mid^2 \,\,\, . \nonumber 
\end{eqnarray} 
Since the term $\mid x\mid^2$ present in (\ref{sumrule}) suppresses 
the ultraviolet singularity of the two-point correlator of 
$\Theta$, the sum rule (\ref{sumrule}) is expected to be saturated 
by the first terms of the series (\ref{cspect}). For the SShG model 
the first approximation to the sum rule (\ref{sumrule}) is given by 
the contributions of the two-particle states 
\EQ
\Delta C^{(2)} = \frac{3}{8\pi^2 m^4} \int_0^{\infty} 
\frac{d\beta}{\cosh^4\beta} \,\left[
\mid F_{bb}^{\Theta}(2\beta)\mid^2 +
\mid F_{ff}^{\Theta}(2\beta)\mid^2 \right] \,\,\,.
\label{twocth}
\EN 
Analogous expression is obtained by a fermionic version of the
$c$--theorem \cite{GMsusy}. 
The numerical data relative to the above integral for different values 
of the coupling constant $\alpha$ is reported in the above table. 
\begin{table}
$$ \begin{array}{|c|c|c|c|}
 \hline
& &  & \\
\alpha 
\,\,\,
& \frac{\lambda^2}{4\pi} \,\,\,
& \Delta\,c^{(2)} \,\,\,
& {\makebox {\rm precision}} $\%$ \,\,\, \\
   &        &   &      \\
\hline
   &        &    &    \\
\frac{1}{100} \,\,\,& \frac{1}{99} \,\,\,& 1.49968 \,\,\,
& 0.0213 \,\,\, \\
\frac{3}{100} \,\,\,& \frac{3}{97} \,\,\,& 1.49741 \,\,\,
& 0.1726 \,\,\,\\
\frac{1}{20}\,\,\,  & \frac{1}{19}\,\,\, & 1.49349 \,\,\,
& 0.4340 \,\,\, \\
\frac{1}{10} \,\,\, & \frac{1}{9}\,\,\,  & 1.47955 \,\,\,
& 1.3633 \,\,\, \\
\frac{3}{20} \,\,\, & \frac{3}{17} \,\,\,& 1.46333 \,\,\,
& 2.4446 \,\,\, \\
\frac{1}{5} \,\,\,  & \frac{1}{4}\,\,\,  & 1.44742 \,\,\,
& 3.5053 \,\,\,\\
\frac{3}{10}\,\,\,  & \frac{3}{7} \,\,\, & 1.42109 \,\,\,
& 5.2606 \,\,\, \\
\frac{2}{5} \,\,\,  & \frac{2}{3} \,\,\, & 1.40480 \,\,\,
& 6.3466 \,\,\, \\
\frac{1}{2}\,\,\,   &  1 \,\,\,          & 1.39935 \,\,\,
 & 6.7100 \,\,\,\\
& &  &\\ \hline
\end{array}
$$
\end{table}
They are 
remarkably close to the theoretical value $\Delta C = \frac{3}{2}$, 
even for the largest possible value of the coupling constant, which 
is the self-dual point $\lambda = \sqrt{4\pi}$. In addition to 
this satisfactory check, a more interesting result is obtained by 
analysing the application of the $c$-theorem sum rule to models 
which are obtained as analytic continuation of the SShG.

For the roaming model, by taking the analytic continuation 
\EQ
\alpha \rightarrow \frac{1}{2} + i \alpha_0 \,\,\,
\label{rrr}
\EN 
and the limit $\alpha_0 \rightarrow \infty$, the result is 
$\Delta C^{(2)} = 0.9924...$, i.e. a saturation within few percent 
of the exact value $\Delta C =1$ relative to this case. 
 
In the analytic continuation $\alpha \rightarrow -\alpha$, 
the $S$-matrix develops a pole in the physical strip located 
at $\beta = 2  \pi\alpha i$. For this model, the operator 
$\Theta(z,\bar z)$ has also a one-particle Form Factor 
$F_{b}^{\Theta}$, which can be easily determined to be 
$
F_b^{\Theta} = 
- 1.6719(3) \,i
$. The series of the sum rule has alternating sign, with the first 
contribution given by the one-particle Form Factor
$
\Delta C^{(1)} = \frac{6}{\pi} (F_b^{\Theta})^2 = -5.3387(4) 
$. This quantity differs for a $1.6\%$ from the theoretical value 
$\Delta C = -\frac{21}{4}=-5.25$. By also including the positive 
contribution of the two-particle FF, computed numerically  
$
\Delta C^{(2)} = 0.09050(8),
$ 
the estimate of the central charge of the model further improves, 
$C= -5.2482(4)$, with a difference from the exact value of just  
$0.033\%$.  

The Form Factors above determined for the trace operator $\Theta(x)$ 
can be used also in this case to estimate its two-point correlation function. 
The graph of this function is shown in Figure 2: note that this function 
diverges at the origin, in agreement with the positive conformal 
dimension $\Delta = 1/4$ of this operator, but it presents a non--monotonous 
behaviour for the alternating sign of its spectral series.

\vspace{-3cm}
\begin{center}
\begin{minipage}[b]{.65\linewidth}
\centering\psfig{figure=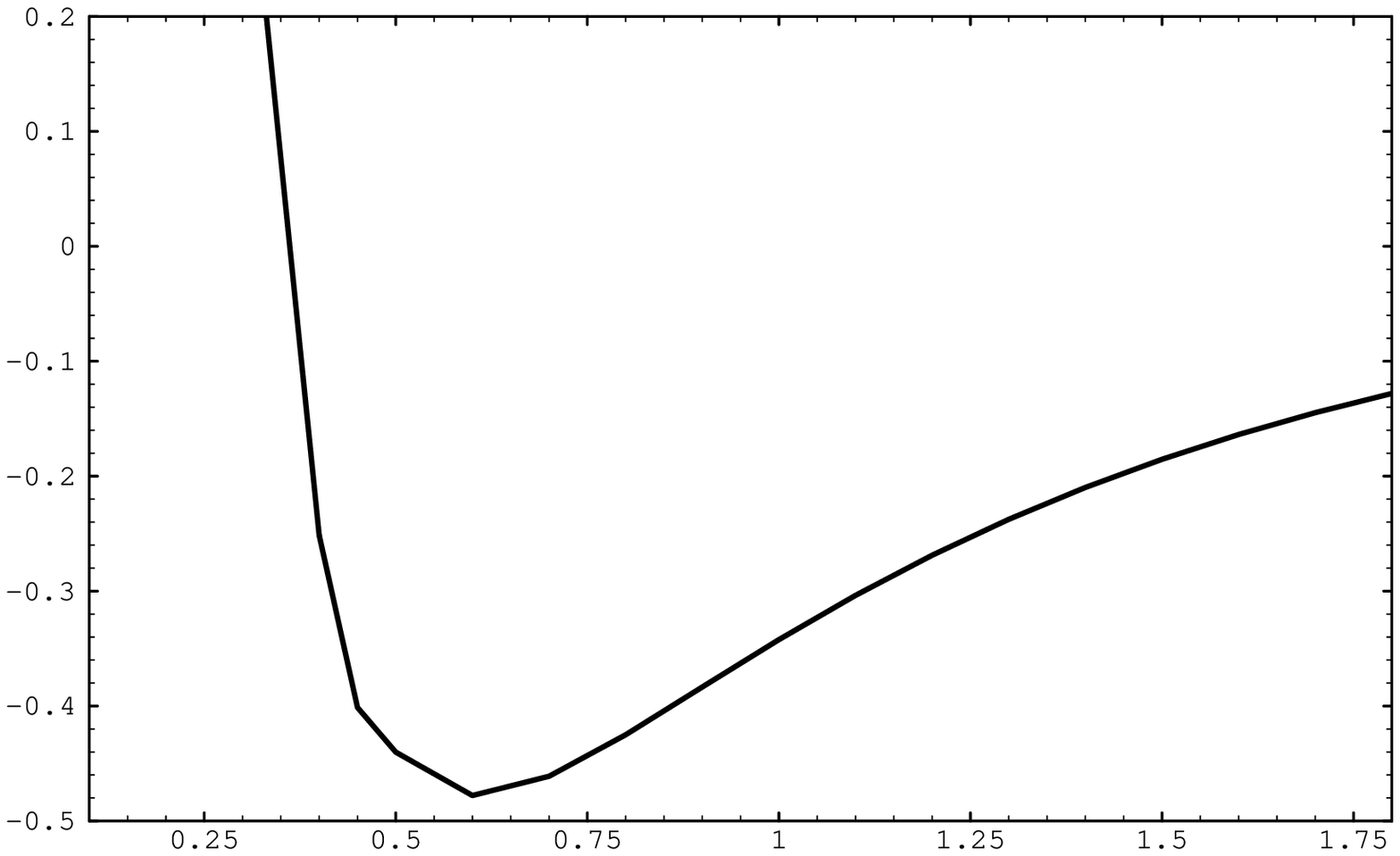,width=\linewidth}

\vspace{-3cm}
\begin{center}
Figure 2. 
\end{center}
\end{minipage} \hfill
\end{center}

\end{document}